# Non-Contact Measurement of Thermal Diffusivity in Ion-Implanted Nuclear Materials

F. Hofmann[1]*, D.R. Mason[2], J.K. Eliason[3], A.A. Maznev[3], K.A. Nelson[3], S.L. Dudarev[2]

[1] Department of Engineering Science, University of Oxford, Parks Road, Oxford, OX1 3PJ, UK

[2] CCFE, Culham Science Centre, Abingdon, OX14 3DB, UK

[3] Department of Chemistry, Massachusetts Institute of Technology, 77 Massachusetts Avenue, Cambridge, MA 02139, USA

*   *felix.hofmann@eng.ox.ac.uk*                             +44 1865 283 446

**Knowledge of mechanical and physical property evolution due to irradiation damage is essential for the development of future fission and fusion reactors. Ion-irradiation provides an excellent proxy for studying irradiation damage, allowing high damage doses without sample activation. Limited ion-penetration-depth means that only few-micron-thick damaged layers are produced. Substantial effort has been devoted to probing the mechanical properties of these thin implanted layers. Yet, whilst key to reactor design, their thermal transport properties remain largely unexplored due to a lack of suitable measurement techniques.**

**Here we demonstrate non-contact thermal diffusivity measurements in ion-implanted tungsten for nuclear fusion armour. Alloying with transmutation elements and the interaction of retained gas with implantation-induced defects both lead to dramatic reductions in thermal diffusivity. These changes are well captured by our modelling approaches. Our observations have important implications for the design of future fusion power plants.**

Nuclear fusion is an ideal sustainable energy source. A major hurdle to its commercial development is the availability of sufficiently resilient materials. Tungsten-based alloys are the main candidates for plasma-facing components in future magnetic confinement fusion reactors[1]. In a demonstration (DEMO) reactor they will be exposed to high temperatures (~1500 K), irradiation with 14.1 MeV fusion neutrons and a large flux of energetic ions (up to 15 MWm$^{-2}$)[2,3]. High thermal conductivity is one of the main material selection criteria[4]. A significant degradation of thermal conductivity could result in excessive temperatures with potentially disastrous consequences for fusion armour integrity[5].

Exposure of fusion armour to 14.1 MeV neutrons leads to cascade damage and transmutation alloying. Calculations indicate that after 5 years of operation, initially pure tungsten (W) in a DEMO divertor would contain up to 4 atomic % rhenium (Re)[6]. A W-5%Re alloy has less than half the room temperature thermal diffusivity of pure tungsten[7,8]. Quantifying the effects of fusion neutron cascade damage on thermal conductivity is more difficult. As a proxy, thermal transport in fission neutron irradiated tungsten has been considered[9,10]. A damage level of 0.6 displacements per atom (dpa), that would be reached in 3 months in DEMO[6], caused a reduction of room temperature thermal conductivity by 25%[10].

An interesting role is played by helium, which is implanted from the plasma into the tungsten matrix. At elevated temperatures helium migrates from surfaces into the bulk, and strongly interacts with irradiation induced defects[11], binding to vacancies[12,13] and supressing their recombination with self-interstitial atoms (SIAs)[14]. Helium-ion implantation is an efficient tool to study this interaction[15] and great effort has been invested into the development of micro-mechanics approaches able to quantify the mechanical properties of micron-thin ion-implanted layers[16-18].

The thermal transport properties of ion-damaged layers, however, are as yet largely unexplored due to a lack of suitable techniques. The references cited above used either a laser flash technique[8-10] or electrical resistivity measurements[7]. Both are only suited to bulk samples. Recently two new approaches, the 3-omega technique[19] and thermal reflectance measurements[20,21], have been proposed to quantify thermal transport in thin, ion-irradiated surface layers. The former required deposition of complex surface features on the sample and showed significant experimental uncertainties. The latter required samples to be coated and the probed depth depended on the, *a priori* unknown, thermal diffusivity.

Here we present a new, entirely different approach to measuring the thermal transport properties of ion-implanted materials. Using the non-contact laser-induced transient grating (TG) technique[22] we quantify thermal diffusivity in few-micron-thin layers of helium-implanted tungsten. The effect of transmutation alloying is mimicked by considering tungsten-rhenium alloys. In both types of samples we find substantial changes in thermal diffusivity. They are analysed using a kinetic theory model, providing insight into the underlying defect distribution. Our results are discussed in the light of current design practise for future fusion reactors.

**Results:**

TG measurements use two short excitation laser pulses that are overlapped on the sample with a well-defined crossing angle (Fig. 1 (a)). Interference of the pulses produces a spatially sinusoidal intensity grating with a fringe spacing $\lambda$[23]. Absorption of the light leads to a temperature grating with period $\lambda$. Rapid thermal expansion also launches two counter-propagating surface acoustic waves (SAWs)[24]. Both the thermal grating and the SAWs cause displacements of the sample surface. These are detected by diffraction of a quasi-continuous probe beam, heterodyned with a reflected reference beam. Fig. 1 (b) presents the signal measured from a pure tungsten sample showing a number of oscillations, due to the propagating SAWs, superimposed on a background due to the decaying temperature grating.

On the surface of bulk samples thermal transport occurs both in-plane, from peaks to troughs of the thermal grating, and into the depth of the sample. The surface profile due to the thermal grating follows a non-exponential decay[23]:

$$\frac{\partial u_z}{\partial x} \propto \mathrm{erfc}\left(q\sqrt{\alpha t}\right). \tag{1}$$

Here $q = 2\pi/\lambda$, $t$ is time and $\alpha$ is the isotropic thermal diffusivity, $\alpha = \kappa/C$, where $\kappa$ is thermal conductivity and $C$ the volumetric heat capacity. Thermal diffusivity is determined by fitting Eqn. 1 to the experimental data, taking into account the sinusoidal variation due to SAWs. Fig. 1 (b) shows the fit to pure tungsten data.

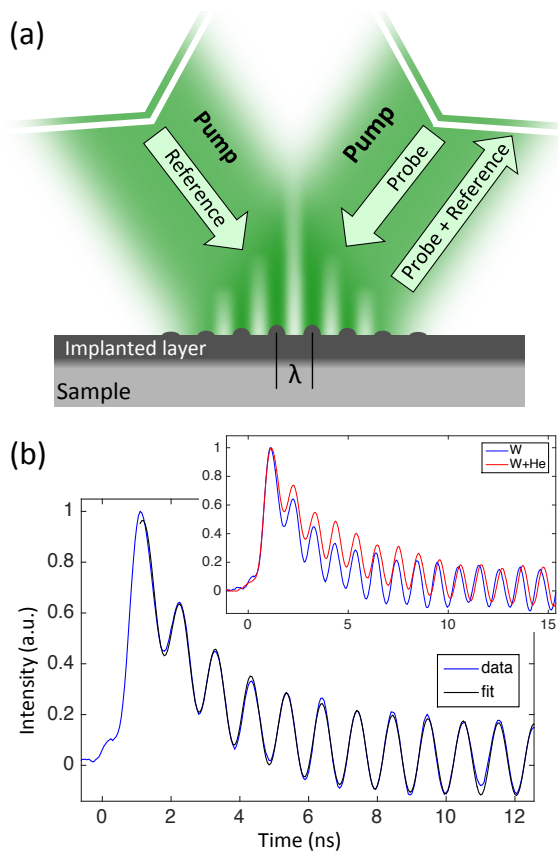

Figure 1: **Transient grating thermal transport measurements** (a) Schematic of the transient grating measurement setup, showing an ion-implanted sample. (b) Experimentally recorded time trace of scattered probe intensity for the pure tungsten sample at 296 K. Also shown is a fit to the data. Inset are time traces recorded for pure tungsten and tungsten implanted with 3100 appm of helium, both at 296 K. Thermal grating decay in the implanted sample is visibly slower.

This method has several important advantages over previously mentioned approaches. No sample coating or indeed contact with samples is required. The probed depth is directly set by the experimental geometry. The accuracy of thermal measurements is intrinsically high since the technique does not rely on measuring a temperature difference or heat flux. Simultaneously the measurements also provide surface acoustic wave data that can be used for very sensitive measurements of elastic properties[24].

Thermal transport measurements were carried out on 99.9% pure tungsten and on tungsten rhenium alloy samples (W-1%Re, W-2%Re). Two pure tungsten samples were helium-implanted at 300°C, using multiple ion energies to achieve approximately uniform predicted helium concentrations of 280 ± 40 atomic parts per million (appm) and 3100 ± 480 appm respectively in a 2.6 µm thick surface layer[25]. The associated implantation damage, predicted using the Stopping Range of Ions in Matter (SRIM) code[26], was 0.017 ± 0.004 dpa and 0.19 ± 0.04 dpa respectively. Implantation profiles are shown in supplementary Fig. 1.

An important question concerns the choice of the transient grating period, $\lambda$. In bulk samples TG measurements probe thermal properties up to a depth of approximately $\lambda/\pi$ [23]. The implanted layer thickness in the helium-implanted samples is approximately 2.6 µm. To ensure that the unimplanted substrate did not affect the thermal transport measurements, a value of $\lambda$ = 2.74 µm was chosen. Pure tungsten was measured at temperatures of 140 K, 200 K, 296 K, 373 K and 473 K, whilst all other samples were measured only at the latter three temperatures. Fig. 2 and Fig. 3 show the measured thermal diffusivities for the alloy samples and the helium-implanted samples respectively.

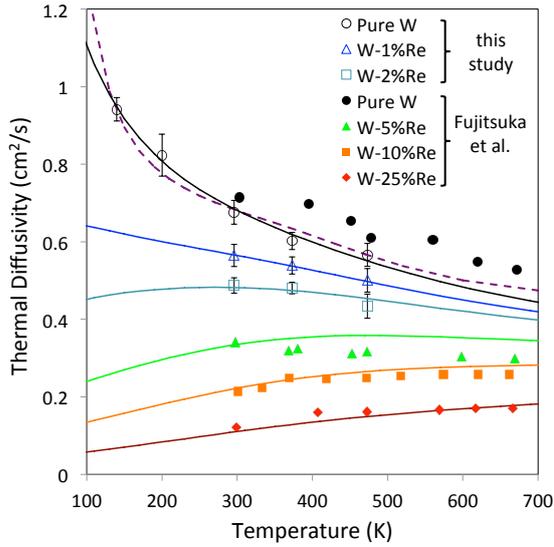

*Figure 2: **Thermal diffusivity of tungsten-rhenium alloys** Measured (open symbols) and modelled (solid lines) thermal diffusivity. Also shown is literature reference data for the thermal diffusivity of pure tungsten[27-29] (dashed purple line). The thermal diffusivity can be extrapolated from the fit to our data and compared to laser-flash measurements reported by Fujitsuka et al.[8] (solid symbols). Variation in the model lines due to uncertainty in the scattering rate estimate is similar to the symbol size.*

**Discussion:**

The measured thermal diffusivity of tungsten alloys can be analysed using a kinetic theory model (see details in supplementary section 3). Provided we remain in the dilute alloy limit, the principal carriers of heat at or above the Debye temperature (312 K in tungsten[28]) are electrons[30]. Variations in thermal diffusivity can then be attributed to changes in the electron scattering time $\tau_e$. According to Matthiessen's rule, the total electron scattering rate will be the sum of rates of scattering from impurities, phonons and other electrons, subject to the Ioffe-Regel limit that the electron mean-free-path cannot be much smaller than the separation between atoms[31,32]. A fit of this model to our experimental data for pure tungsten is shown in Fig. 2 (black line). Our measurements are in very good agreement with literature reference data for pure tungsten (Fig. 2 dashed line)[27-29] and lie within less than 10% of the data measured by Fujitsuka et al.[8] (also plotted in Fig. 2). This provides conclusive evidence of the reliability and accuracy of the TG technique for the contact-less characterisation of thermal transport.

For the tungsten-rhenium alloys, we fit a further parameter, the scattering rate due to a rhenium atom in a tungsten matrix, finding $\sigma_{0,Re}$ = 1.38±0.1 THz. We can then extrapolate to higher rhenium concentrations. Shown in Fig. 2 are lines for 5% to 25% rhenium alloys, with corresponding experimental measurements by Fujitsuka et al[8]. Even though these rhenium concentrations are beyond the expected limit of validity of the dilute-alloy model, the match is surprisingly good, demonstrating that the TG technique delivers reliable and transferable thermal parameters.

At low temperatures, using the Wiedemann-Franz law[30], we can estimate that the change in resistivity due to a single rhenium atom in the tungsten matrix is $\delta\rho_{Re}$ = 127±10 μΩcm/atomic fraction. This value can be compared to the directly measured experimental value $\delta\rho_{Re}$ = 145 μΩcm/ atomic fraction[7]. Clearly the agreement is good, although our estimate is a little low as we neglected phonon conductivity in our treatment.

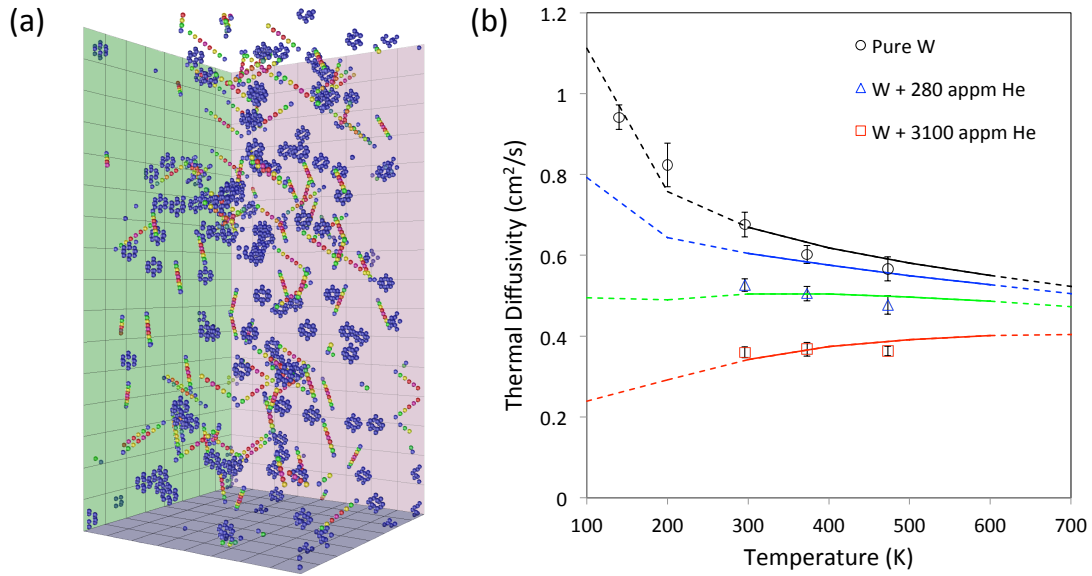

*Figure 3: **Thermal diffusivity of helium-implanted tungsten** (a) Electron scattering rates on high energy atom sites coloured from blue (0.5 THz) to pink (3.5 THz). 900 appm Frenkel pair defects in 131,000 tungsten atoms at 300 K. Vacancies are seen as blue "cages" of 8 atoms surrounding the vacant site, while interstitials are seen as ½ <111> crowdions. Bulk atoms are not shown. (b) Measured (open symbols) and modelled (lines) thermal diffusivity for helium-implanted tungsten. The model lines are for 0 (black), 300 (blue), 900 (green) and 3000 (red) appm Frenkel pairs in bulk tungsten. Dotted lines mark extrapolation outside the fitted temperature region.*

In the helium-implanted tungsten samples damage will be present in the form of vacancies and self-interstitials[33]. As conducting electrons can be scattered from any atom associated with a vacancy or interstitial defect, it is inappropriate to treat these defects as very strong point scatterers. Instead we use an empirical atomistic model to estimate thermal conductivity for the damaged layer[32]. This model computes electron-phonon scattering locally at each atomic site and the larger contribution from Mott-Jones impurity scattering at atoms with significantly more energy than the thermal average. The total scattering due to a helium-filled vacancy is likely to be very similar to that for an empty monovacancy, as the helium atoms do not contribute valence electrons[34]. Hence this atomistic model is fitted only to the measured conductivity of pure tungsten in the range 300 - 600 K[28] and the resistivity per Frenkel pair[35]. The scattering rate on atomic sites is shown in Fig. 3 (a), highlighting the distributed nature of scattering due to vacancy and interstitial defects.

At the implantation temperature of 573 K, vacancy mobility is low and no significant vacancy clustering and bubble growth is expected[36,37]. However, it is difficult to exactly quantify the number of interstitials and vacancies retained in the sample after implantation at finite temperature. Damage calculations using SRIM[26] indicate that approximately 60 Frenkel pairs are formed per injected helium, most of which recombine almost immediately after generation. The changes we observe are due to the few residual defects that have not recombined. Therefore we modelled thermal diffusivity for a range of Frenkel pair concentrations (Fig. 3 (b)). Curves predicted for 900 appm and 3000 appm Frenkel pairs provide a good fit to the experimental data for low (280 appm) and high helium dose (3100 appm) implanted samples respectively. The helium to vacancy ratio of ~1:3 at the lower implantation dose can be compared to the value of 1:5 calculated by Becquart[38], who pointed out that at low helium doses defect retention is dominated by impurities, most notably carbon. In the present tungsten samples carbon is present at a concentration of ~900 appm. At the higher implantation dose the helium to vacancy ratio approaches 1:1 as the relative importance of impurities decreases and Frenkel pair recombination is predominantly hindered by helium occupying vacancies[24].

The measured room temperature thermal diffusivity in the high helium dose sample (0.36 $cm^2s^{-1}$) can also be compared to that of other samples with similar levels of cascade damage (~0.2 dpa). In neutron irradiated tungsten (623 K, 0.2 dpa) thermal diffusivity was much higher (0.60 $cm^2s^{-1}$)[9], whilst in Cu-ion implanted tungsten (298 K, 0.2 dpa), a significantly lower value was measured (0.29 $cm^2s^{-1}$)[19]. This suggests that the calculated dpa does not provide a reliable guide to the change in thermal properties. This is particularly the case since retained gas and impurities clearly play a central role in determining defect removal/annihilation rates.

In conclusion we have demonstrated the feasibility of high fidelity thermal transport measurements in micron-thick ion-implantation-damaged surface layers. Our approach opens the door to a comprehensive characterisation of thermal transport properties in ion-implanted materials for nuclear materials research. The transient grating technique requires no physical contact, making it ideal for the characterisation of active samples and *in situ* measurements. We have shown that the change in thermal diffusivity is not a trivial function of ion fluence and can offer an insight into the post-implantation microstructural evolution of defects invisible in transmission electron microscopy. At present the degradation of thermal transport properties due to irradiation damage is not taken into account in the design of DEMO divertor armour[39,40]. Given the substantial reductions in thermal diffusivities we have measured, it seems essential that these effects be considered in future design iterations.

**Acknowledgements**

We are grateful to C. Beck for preparing the samples and N. Peng for carrying out the ion implantation. FH acknowledges funding from the John Fell fund (122/643) and the Royal Society (RG130308). Transient grating measurements at MIT were supported as part of the S3TEC Energy Frontier Research Center funded by the U.S. Department of Energy, Office of Basic Energy Sciences under award no. DE-SC0001299/DE-FG02-09ER46577. In part this work has been carried out within the framework of the EUROfusion Consortium and has received funding from the Euratom research and training programme 2014-2018 under grant agreement no. 633053. To obtain further information on the data and models underlying this paper please contact PublicationsManager@ccfe.ac.uk. The views and opinions expressed herein do not necessarily reflect those of the European Commission. This work was part-funded by the United Kingdom Engineering and Physical Sciences Research Council via programme grants EP/G050031 and EP/H018921.

**Methods**

Sample preparation

Tungsten and tungsten-rhenium samples were produced by arc melting of high purity elemental powders (99.9%). The resulting slugs were sectioned and mechanically polished, finishing with a colloidal silica polishing step. Optical microscopy showed grains up to 1 mm in size. Electron-backscatter diffraction showed no significant texture. Tungsten samples were helium-ion-implanted at the National Ion Beam Centre, University of Surrey, UK using 12 different ion energies from 0.05 MeV to 1.8 MeV[25]. Implantation and collision damage profiles (supplementary Fig. 1) were calculated using the Stopping Range of Ions in Matter code[26] with a tungsten displacement energy of 68 eV.

Transient grating measurements

Two short excitation pulses (515 nm wavelength, 60 ps pulse duration and 1.75 µJ pulse energy) were used to generate the transient grating. Most measurements were performed using a grating period $\lambda$ = 2.74 µm, with some confirmation measurements in pure tungsten using $\lambda$ = 9 µm also. Deviation of the actual grating period from the nominal value was negligible (less than 0.5%), confirmed by comparing the measured surface acoustic wave velocity in tungsten to the literature value[24]. Temperature grating decay and surface acoustic waves were probed by diffraction of a quasi-continuous probe beam (532 nm wavelength, 10 mW average power). The diffracted beam was heterodyned with a reflected reference beam (Fig. 2 (a)) and the combined beam directed to a fast avalanche photo-diode. Time traces were recorded on an oscilloscope. The bandwidth of the detection system was approximately 2 GHz. Excitation and probe spot sizes of 500 µm and 150 µm diameter respectively at $1/e^2$ intensity level were used. All measurements were carried out in a cryostat under medium vacuum ($10^{-2}$ mbar).